\documentclass{article}

\usepackage{arxiv}

\usepackage[utf8]{inputenc} 
\usepackage[T1]{fontenc}    
\usepackage{hyperref}       
\usepackage{url}            
\usepackage{booktabs}       
\usepackage{amsfonts}       
\usepackage{nicefrac}       
\usepackage{microtype}      
\usepackage{lipsum}
\usepackage{indentfirst}
\usepackage{setspace}

\usepackage{graphics}      
\usepackage{graphicx}      
\usepackage{longtable}     
\usepackage{float}

\newcommand{\beq}{\begin{equation}}
\newcommand{\eeq}{\end{equation}}
\newcommand{\bea}{\begin{eqnarray}}
\newcommand{\eea}{\end{eqnarray}}

\title{CASIMIR EFFECT NEARBY AND THROUGH A COSMOLOGICAL WORMHOLE}

\author{{\Large
  A. C. L. Santos,\thanks{alana.santos@aluno.uece.br}\,\,\,\, C. R. Muniz\thanks{celio.muniz@uece.br}\,\,\,\, and\,\,\, L. T. Oliveira\thanks{leonardo.tavares@uece.br}}\\
{\Large Universidade Estadual do Cear\'{a}}\\
{\Large Faculdade de Educa\c{c}\~{a}o, Ci\^{e}ncias e Letras de Iguatu}\\
{\Large Av. Dario Rabelo, s/n,
CEP: 63.500-000, Iguatu, CE, Brazil}\\
}

\begin{document}
\maketitle
\vspace{1.0cm}
\doublespacing 
\begin{abstract}
{\large 
In this letter, we investigate the changes in the quantum vacuum energy density of a massless scalar field inside a Casimir cavity that orbits a wormhole, by considering the cosmological model with an isotropic form of the Morris-Thorne wormhole, embedded in the FLRW universe. In this sense, we examine the effects of its global curvature and scale factor in an instant of the cosmic history, besides the influences of the local geometry as well as of inertial forces, on the Casimir energy density. We also study the behavior of this quantity when each plate is fixed  without rotation at the opposite sides of the wormhole throat, at zero and finite temperatures, taking into account the effective distance between the plates through the wormhole throat. 
}

\vspace{0.5cm}
\noindent{{\large Keywords: Cosmological wormhole. Gravito-inertial effects. Casimir energy.}}
\end{abstract}

\renewcommand{\thesection}      {\Roman{section}}
\maketitle
\newpage
{\large
\indent Wormholes are solutions to the Einstein's equations of General Relativity and represent physical connections between two distant regions of the universe \cite{Visser}. Although the first solutions had obstacles with respect to the traversing through them \cite {EinsteineRose, Wheeler, Ker1, Ker2, Ker3}, Michael Morris and Kip Thorne investigate a viable traversable wormhole, under certain specific conditions \cite{KipThorne}.\\
\indent In general, it is necessary some form of exotic matter, {\it i.e.}, the one that obeys non-trivial equations of state, around a wormhole, with some exceptions \cite{wormholecasimir1, wormholecasimir2, wormholecasimir3, wormholecasimir4}. Thus, the Casimir effect, that in its original presentation involves negative energies of quantum fields present between parallel and uncharged plates, has been increasingly examined near such objects. A pioneering analysis by Sorge \cite{Sorge} investigated the interference of both non-inertial effects and spacetime geometry on the vacuum energy density of a non-massive scalar field present in a small Casimir cavity, which orbits an Ellis-Thorne wormhole \cite{Ellis}. In this perspective, recent works considering the Casimir effect around other kinds of wormhole have been published \cite{Alana,Bezerra}, as well as others analyzing the modelling of wormhole shapes via Casimir energy \cite{Remo,Jusufi}. Thus, the study of this effect in various scenarios where gravity plays a relevant role contributes to one understanding how it occurs the fundamental interplay between the vacuum of quantum fields and gravitation, taking more a step towards a final theory.\\
\indent Another deep connection between wormholes and quantum mechanics, further expanding the interest around these objects, was described in \cite{Maldacena} in the so-called conjecture \emph{ER = EPR}. This latter postulates that there is a kind of equivalence between a wormhole like the Einstein-Rose bridge \cite {EinsteineRose} and quantum entanglement, firstly suggested by Einstein-Podolsky-Rose \cite{EPR}. Basically, if there is such a bridge, the two extremities are entangled by EPR pairs, and vice-versa. On the other hand, a counterexample of this relationship was given in \cite{counterexample}, by taking into account pairs of black holes, spatially separated and quantum entangled, in an anti-de Sitter space, preventing the generality of that connection.\\
\indent Recently, Kim \cite{Kim} studied a cosmological model consistent with an isotropic form of the Morris-Thorne wormhole associated with the Friedmann-Lemaître-Robertson-Walker (FLRW) universe given in \cite{Mirza}, finding the exact solution that satisfies Einstein's field equation. That author discussed if the wormhole must interact, in some way, with the exotic matter present in spacetime and with this latter itself \cite{Kim}.\\
\indent In this work, we will investigate the changes in the quantum vacuum energy density of a massless scalar field inside a Casimir apparatus that orbits a wormhole, according to the techniques used in \cite{Sorge,Alana}, but now by considering the cosmological model with an isotropic form of the Morris-Thorne wormhole, embedded in the universe (FLRW), found in \cite{Kim}. Thus, we generalize the work of Sorge \cite{Sorge}, by examining the effects of the Universe global curvature and its scale factor, besides the influences of the local geometry as well as of inertial forces, on the Casimir energy density. We also will study the behavior of such a quantity when each plate is fixed statically at the opposite sides of the wormhole throat, at zero and finite temperatures.\\
\indent Initially, we take the FLRW metric of a cosmological Morris-Thorne wormhole, in isotropic coordinates, described by \cite{Kim} ($G=c=k_B=1$)
\begin{equation}
\displaystyle ds^2 = - dt^2 + \frac{a^2(t)}{(kr^2 +1)^2} \left(1 + \frac{b_0^2}{4r^2}\right)^2[dr^2 + r^2 (d\theta^2 + \sin^2\theta d\phi^2)],   \label{eq1}
\end{equation} 
where $a(t)$ is the scale factor, $k$ the curvature of the Universe and $\displaystyle b_0/2$ the radius of the wormhole throat. Hence, we define the function $C(r)$ from the equation (\ref {eq1}) as
\begin{equation}
\displaystyle C(r)\equiv   \frac{a(t_0)}{(kr^2 +1)}\left(1 + \frac{b_0^2}{4r^2}\right),\label{eq2}
\end{equation}
where we take the scale factor in a specific instant ($t_0$) of the Universe, supposing thus that it is static in a first approximation.\\ 
\indent Consider the Casimir apparatus orbiting, in a circular path, around the wormhole described by Eq. (\ref {eq1}) at the equatorial plane $(\theta=\pi/2)$. Based on what was presented in \cite{Sorge,Alana}, we will adopt the necessary conditions to determine the Casimir energy density in a cosmological scenario. First, we introduce a unitary tangent timelike vector $\displaystyle \mathbf{u}=e^\psi(\partial_t+\Omega\partial_\phi)$, with angular velocity of the plates $\Omega=d\phi/dt$ and, consequently,
\begin{equation}
\displaystyle e^\psi = \frac{1}{\sqrt{1-r^2\Omega^2 C^2(r)}},   
\end{equation}
for which, taking into account the positive direction of the rotation, we have
\begin{equation}
\displaystyle 0\leq \Omega<\displaystyle
 \frac{(kr^2 + 1)}{a(t)r}{\left(1+\frac{b_0^2}{4r^2}\right)^{-2}} \equiv \Omega(r).
\end{equation}

We will consider now the quantum vacuum fluctuations of a massless scalar field $\varphi(x^{\mu})$ confined within the orbiting cavity. Thus, we must initially solve the Klein–Gordon equation, assuming Dirichlet boundary conditions on the plates, whose proper separation, in the comoving observer’s frame, is $L$. The proper area of each plate is $S$ and we will work with the approximation $L\ll\sqrt{S} \ll b_0 \leq r$. In this approach we do not consider tidal effects inside the cavity, only the ones of gravito-inertial nature. We must still mention that there are in the literature controversies about the existence or not of these effects between the plates, at least in lower orders of approximation \cite{Geova,Sorge2}.\\
\indent  Implementing a frame associated to the orbiting Casimir apparatus, so that the azimuth angle transforms as $d\phi\to d\phi+\Omega dt$, the comoving observer will obtain a metric given by
\begin{equation}
\displaystyle ds^2=\left[1-r^2\Omega^2 C^2(r)\right] dt^2- C^2(r)(dr^2+r^2d\theta^2-r^2d\phi^2-2\Omega r^2d\phi dt).
\end{equation}
Introducing orthonormal tetrades in the form of
\begin{eqnarray}
    &{\bf \hat{e}}_{\tau}&=[1-r^2\Omega^2 C^2(r)]^{-1/2}\frac{\partial}{\partial t},\nonumber\\
    &{\bf \hat{e}}_{x}&=C(r)^{-1}\frac{\partial}{\partial r},\nonumber\\
    &{\bf \hat{e}}_{y}&=[r C(r)]^{-1}\frac{\partial}{\partial \theta},\nonumber\\
    &{\bf \hat{e}}_{z}&=r\Omega C(r)[1-r^2\Omega^2 C^2(r)]^{-1/2}\frac{\partial}{\partial t}+[r C(r)]^{-1}[1-r^2\Omega^2 C^2(r)]^{1/2}\frac{\partial}{\partial \phi},
\end{eqnarray}
where $z$ is a coordinate normal to the plates and ${\bf \hat{e}}_{\tau}$ the four-velocity. Considering the minimal coupling, the motion equation for the non-massive scalar field is, therefore
\begin{equation}\label{K-G}
 \displaystyle   \frac{1}{\sqrt{-g}}\partial_{\mu}(\sqrt{-g}g^{\mu\nu}\partial_{\nu}\varphi)=\nabla^2\varphi-\frac{2}{r}\frac{\partial \varphi}{\partial x}=0,
\end{equation}
where $\nabla^2=\partial^2_x+\partial^2_y+\partial^2_z$. We have used the approximation in which $r\approx$ constant inside the plates, provided $L\ll r$. \\
\indent The solutions of Eq. (\ref{K-G}) satisfying the Dirichlet boundary conditions are given by
\begin{equation}
\varphi_{n,{\bf k_{\|}}}=N_n\exp{({-i\omega_{n,{\bf k_{\|}}}}\tau)}\exp{(i{\bf k_{\|}}}\cdot {\bf x_{\|}})\sin{\left(\frac{n\pi z}{L}\right)},
\end{equation}
where ($\omega_{n,\|},{\bf k_{\|}})$ are the eigenfrequencies and momenta of the field free propagation modes parallel to the plates. These solutions are in the flat spacetime form, since tidal effects due to spacetime inhomogeneities inside the small cavity are negligible. These modes are normalized from the Klein-Gordon scalar product, given in the tetrade frame by \cite{Sorge}
\begin{equation}
\displaystyle \langle \varphi_n (\textbf{k}), \varphi_m(\textbf{k}')\rangle = i \int_\sum [(\partial_a \varphi_n){\varphi^*_m} - \varphi_n(\partial_a{\varphi^*_m})]n^a dxdydz, 
\end{equation}
where $n^a = e^a_\mu n^\mu$ and $n^\mu = (1, 0, 0, - \Omega)$. Taking into account that
\begin{equation}
\displaystyle\langle \varphi_n(\textbf{k}_\parallel), \varphi_m(\textbf{k}'_\parallel) \rangle = \delta^2 (\textbf{k}_\parallel - \textbf{k}'_\parallel) \delta_{mn},    
\end{equation}
we arrive at the normalization parameter
\begin{equation}
    N_n=\left(\frac{\sqrt{1-r^2\Omega^2F^2(r)}}{4\pi^2L\omega_{n,\|}}\right)^{1/2},
\end{equation}
 which guarantees the orthonormalization of the field modes and encodes the properties of the considered spacetime.\\
\indent The Casimir energy will be obtained from the regularization of the expected value of the quantum vacuum fluctuations energy, given by \begin{equation} \label{MeanEnergy}
    \langle \epsilon \rangle=\frac{1}{V_p}\int_{\Sigma}d^3{\bf x}\sqrt{g_{\Sigma}} \sum_n\int d^2{\bf k_{\|}}T_{00},
\end{equation}
where the first integration is realized in the Casimir cavity, which has proper volume $V_p=V \sqrt{-g_{\Sigma}}$, with $V$ being the  volume measured by a distant observer, $g_{\Sigma}=\det{(\hat{g}_{\mu\nu})/\hat{g}_{tt}}$ \cite{Zhang}, and the second integration is over the space of the momenta parallel to the plates. The purely temporal component of the energy-momentum tensor, $T_{00}$, associated to the $n$ mode, is given by
\begin{equation}\label{MomentEnergy}
    T_{00}=\partial_{\tau}\varphi_n\partial_{\tau}\varphi_n^{*}-\frac{1}{2}\eta_{00}\eta^{ij}\partial_i\varphi_n\partial_j\varphi_n^{*}.
\end{equation} 
Plugging (\ref{MomentEnergy}) into (\ref{MeanEnergy}), we obtain
\begin{equation}
    \langle \epsilon \rangle=\frac{\sqrt{1-r^2\Omega^2C^2(r)}}{8\pi^2L}\sum_n\int_0^{\infty} d^2\mathbf{k}_{{\|}}\sqrt{k^2_{\|}+\frac{n^2\pi^2}{L^2}}.\end{equation}
From the Schwinger proper-time representation for the above integral, given by
\begin{equation}
    a^{-z}=\frac{1}{\Gamma(z)}\int_0^{\infty}t^{z-1}\exp{(-a t)}dt,
\end{equation}
in which $a=k^2+n^2\pi^2/L^2$ and $z=-1/2$, with $k_{\|}=k$ and $d^2k_{\|}=2\pi k dk$. Thus, the integral in the momentum variable can be performed by using the Euler representation for the gamma function, and the summation in $n$ is carried by means of the definition of the Riemann zeta function, $\zeta(s)=\sum_1^{\infty}n^{-s}$. With this, we finally arrive at the Casimir energy density between the plates, given by
\begin{equation}\label{CasEn}
 \displaystyle \epsilon_C =-\left[{1-r^2\Omega^2\frac{a^2(t_0)}{(kr^2 +1)^2}\left(1+\frac{b_0^2}{4r^2}\right)^2}\right]^{1/2}|\epsilon_0|,
\end{equation}
where the factor $\epsilon_0$ is the Casimir energy density of the scalar field in the Minkowsky spacetime, namely
\begin{equation}\label{CasMink}
\epsilon_0 = -\frac{\pi^2}{1440L^4}.
\end{equation}
It is worth to notice that the quantity in Eq.(\ref{CasEn}) reduces to the one found in \cite{Sorge} for $a(t_0)=1$ and $k=0$.\\ 
\indent Looking at the graph of Fig. 1, we can see that the Casimir energy density has its highest values in the space with negative global curvature (hyperbolic), followed by an intermediary energy in the space of zero curvature, and lower energy when the space is positively curved (spherical).
\begin{figure}[H]
\centering
\includegraphics[scale=0.7]{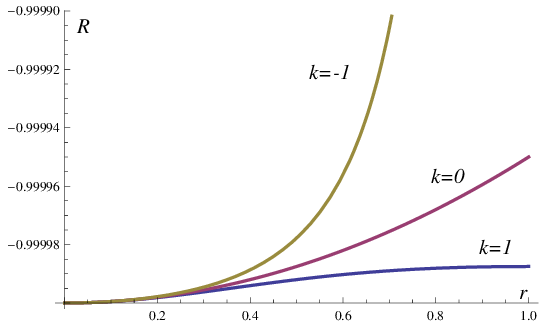}
\caption{The ratio $R=\epsilon_{C}/|\epsilon_0|$, as a function of the radial coordinate, $r>b_0/2$, for $b_0=0.001$, $\Omega=0.01$, and $a(t_0)=1$.}
\end{figure} 
It is interesting to examine the situation in which each plate stays at rest ({\it i.e.}, $\Omega=0$) at fixed radial coordinates $\pm r$. The opposite signs indicate that each plate is positioned with respect to the center of wormhole throat at respective coordinate $r$ in opposite sides. Then, the effective separation between the plates through the wormhole will be given by \cite{Morris}
\begin{equation}\label{Leff}
    L_{eff}=\int_{-r}^{+r}\frac{a(t_0)}{(kr'^2 +1)} \left(1 + \frac{b_0^2}{4r'^2}\right) dr'=a(t_0)\left[\frac{(4-b_0^2 k) \tan^{-1}\left(\sqrt{k} r\right)}{2 \sqrt{k}}-\frac{b_0^2}{2 r}\right].
\end{equation}
We notice that when the plates are nearby the throat radius, $r\approx b_0/2$, the effective distance between them will only have a finite value in a hyperbolic Universe. For a throat very small compared to the Universe curvature radius, $R_U$, we will get 
\begin{equation}\label{LEFF}
    L_{eff}\approx \frac{a(t_0)b_0^3}{3R_U^2}, 
\end{equation}
and the Casimir energy density (\ref{CasMink}) becomes
\begin{equation}
\displaystyle \epsilon_C\approx -\frac{9\pi^2R_U^8}{160a^4(t_0)b_0^{12}},
\end{equation}
which can represent a huge quantity of energy.\\
\indent In which follows, we will made the analysis of the thermal Casimir effect. According to \cite{Zhang}, the present problem reduces to that one of the flat spacetime, since the renormalized Helmholtz free energy (RHFE) associated to the Casimir apparatus orbiting the wormhole in a thermal bath has the same value than the one measured in the Minkowsky spacetime, relative to a comoving observer. Thus, our discussion becomes interesting only when we consider the plates once more fixed at the opposite sides of the wormhole throat. In this context, the proper thermal correction to RHFE is given by \cite{Zhang}
\begin{equation}\label{FreeEn}
   \Delta_{T}F= -\frac{S}{32\pi L_{eff}^3}\large{\sum_{n=1}^{\infty}\left[\frac{\coth{(n\pi\tilde{\beta)}}}{(n\tilde{\beta})^3}+\frac{\pi}{(n\tilde{\beta})^2\sinh^2{(n\pi\tilde{\beta})}}\right]+\frac{\pi^2SL_{eff}T^4}{90}\large},
\end{equation}
where $\tilde{\beta}=1/2TL_{eff}$, with $L_{eff}$ given by (\ref{Leff}).\\
\indent Next, we consider the renormalized thermal correction to the Casimir energy - the proper thermal internal energy - which is defined as
\begin{eqnarray}\label{30}
U^{ren}(T)=-T^{2}\frac{\partial}{\partial T}\bigg(\frac{\Delta_TF}{T}\bigg).
\end{eqnarray}
From Eq. (\ref{FreeEn}), one obtains
\begin{eqnarray}\label{31}
\displaystyle U^{ren}(T)=\frac{S}{16\pi L^{3}_{eff}}\bigg \{\sum^{\infty}_{m=1}\bigg[\frac{\coth(\pi m\tilde{\beta})}{(m\tilde{\beta})^{3}}
+\frac{\pi}{(m\tilde{\beta})^{2}\sinh^{2}(\pi m\tilde{\beta})}+\frac{\pi^{2}\coth(\pi m\tilde{\beta})}{m\tilde{\beta}\sinh^{2}(\pi m\tilde{\beta})}\bigg]
-\frac{\pi^{3}}{30\tilde{\beta}^{4}}\bigg \}.\nonumber \\
\end{eqnarray}
We notice that this expression depends on the proper temperature and the proper geometrical parameters, including the effective distance between the plates through the wormhole.\\
\indent Here, we consider again the Casimir thermal energy density, $u^{ren}=U^{ren}/SL_{eff}$, in the limit $b_0/R_U\ll 1$ in a hyperbolic Universe, which is given by
\begin{equation}\label{38}
u^{ren}(T)\approx\ \frac{3\zeta(3)R_U T^{3}}{2\pi a(t_0)b_0^3}-\frac{\pi^{2}T^{4}}{30}+\frac{27\pi R_U^6 T}{2a^3(t_0)b_0^3}e^{-3\pi R_U/[3a(t_0)b_0T]},
\end{equation}
where we taken into account Eq. (\ref{LEFF}). The leading term is the first one, and the last is a little exponential  correction to Eq.~(\ref{38}), for $m=1$. From Eq. (\ref{38}) we can find that the Casimir thermal energy goes to zero when the temperature vanishes. In Fig.2 we depict the Casimir thermal energy density as a function of the ratio $b=b_0/2R_U\ll 1$, and the temperature, $T$. 
\begin{figure}[!h]
\centering
\includegraphics[scale=1.16]{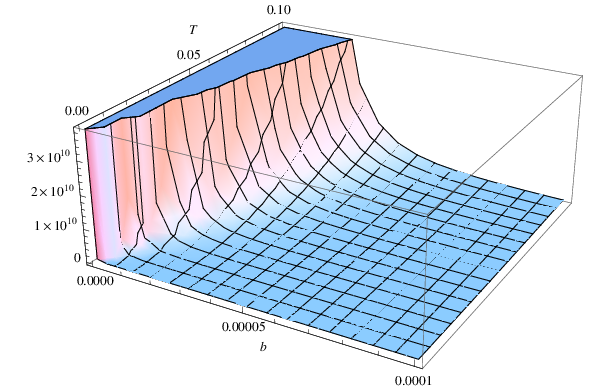}
\caption{Casimir thermal energy density, as a function of the ratio, $b=b_0/2R_U$, and proper temperature, $T$,  for $a(t_0)=1$.}
\end{figure}\\
\indent In the present letter we have investigated the changes in the quantum vacuum energy density of a massless scalar field inside a Casimir apparatus that orbits a wormhole, according to the approach found in \cite{Sorge,Alana}. We have considered the cosmological model with an isotropic form of the Morris-Thorne wormhole, embedded in the universe (FLRW), found in \cite{Kim}. In this sense, we generalize the work of Sorge \cite{Sorge}, by examining the effects of the Universe global curvature, $k$, and its scale factor in a specific instant of the Universe history, $a(t_0)$, beyond the influences of the local geometry as well as of inertial forces, on the Casimir energy density. We found that for a flat Universe $k=0$ and $a(t_0)=1$, this quantity is the same obtained in \cite{Sorge}. We also have found that the Casimir energy density is higher in a hyperbolic Universe, lower in a spherical one, and intermediary in a flat Universe, with the difference between them being higher as more distant the plates are from the wormhole throat. It is worth also point out that their absolute magnitude is always lower than that one measured in the Minkowsky spacetime with the plates at rest.

Finally, we have studied the behavior of this quantity when each plate is fixed without rotation at opposite sides of the wormhole throat, at zero and finite temperatures. In this scenario, we have taken into account the effective distance between them through the wormhole throat. We have shown that the Casimir energy density and its thermal counterpart only are finite in a hyperbolic Universe, when the plates are on the throat radius, and that these energies can be considerably large for short radii compared with the Universe curvature radius, since the effective distance between the plates is much shorter.  

CRM would like to thank Conselho Nacional de Desenvolvimento Cient\'{i}fico e Tecnol\'{o}gico (CNPq) and Fundação Cearense de Apoio ao Desenvolvimento Científico e Tecnológico (FUNCAP), under grant PRONEM PNE-0112-00085.01.00/16, for the partial financial support.

\bibliographystyle{unsrt}  


}

\end{document}